%% file: paper284_final.tex
\documentstyle[10pt,epsf,epsfig,hangcaption,xspace,amssymb,amsfonts,amsmath,amsthm,cite,
dp_delphititle,lineno]{dp_delphi}
\makeindex
\def\DpPaperGroup{EP}
\def\DpPaperRef{2003-035}
\def\DpDate{22 Mai 2003}
\def\DpAuthors{DELPHI Collaboration}
\def\DpTitle{{
The \boldmath $\eta_{c}(2980)$ formation in 
two-photon collisions at LEP energies}}
\def\DpSubmit{(Accepted by Eur. Phys. J. C)}
\def\DpComment{  }
\def\DpEMail{  }

\include{newcom}

\newcommand{\mumu}{\ifmmode {\mu^+\mu^-} \else ${\mu^+\mu^-} $ \fi}

\newcommand{\ba}{\begin{array}}
\newcommand{\ea}{\end{array}}
\newcommand{\bc}{\begin{center}}
\newcommand{\ec}{\end{center}}
\newcommand{\be}{\begin{equation}}
\newcommand{\enq}{\end{equation}}
\newcommand{\beq}{\begin{eqnarray}}
\newcommand{\eeq}{\end{eqnarray}}
\newcommand{\bes}{\begin{eqnarray*}}
\newcommand{\ees}{\end{eqnarray*}}
\newcommand{\Kz}{\ifmmode {\rm K^0_s} \else ${\rm K^0_s} $ \fi}
\newcommand{\Zz}{\ifmmode {\rm Z^0} \else ${\rm Z^0 } $ \fi}
\newcommand{\qqbar}{\ifmmode {\rm q\bar{q}} \else ${\rm q\bar{q}} $ \fi}
\newcommand{\ccbar}{\ifmmode {\rm c\overline{c}} \else ${\rm c\overline{c}} $ \fi}
\newcommand{\bbbar}{\ifmmode {\rm b\overline{b}} \else ${\rm b\overline{b}} $ \fi}
\newcommand{\xxbar}{\ifmmode {\rm x\bar{x}} \else ${\rm x\bar{x}} $ \fi}
\newcommand{\rphi}{\ifmmode {\rm R\phi} \else ${\rm R\phi} $ \fi}
\begin{document}
\makeatletter
\newcount\@tempcntc
\def\@citex[#1]#2{\if@filesw\immediate\write\@auxout{\string\citation{#2}}\fi
  \@tempcnta\z@\@tempcntb\m@ne\def\@citea{}\@cite{\@for\@citeb:=#2\do
    {\@ifundefined
       {b@\@citeb}{\@citeo\@tempcntb\m@ne\@citea\def\@citea{,}{\bf ?}\@warning
       {Citation `\@citeb' on page \thepage \space undefined}}%
    {\setbox\z@\hbox{\global\@tempcntc0\csname b@\@citeb\endcsname\relax}%
     \ifnum\@tempcntc=\z@ \@citeo\@tempcntb\m@ne
       \@citea\def\@citea{,}\hbox{\csname b@\@citeb\endcsname}%
     \else
      \advance\@tempcntb\@ne
      \ifnum\@tempcntb=\@tempcntc
      \else\advance\@tempcntb\m@ne\@citeo
      \@tempcnta\@tempcntc\@tempcntb\@tempcntc\fi\fi}}\@citeo}{#1}}
\def\@citeo{\ifnum\@tempcnta>\@tempcntb\else\@citea\def\@citea{,}%
  \ifnum\@tempcnta=\@tempcntb\the\@tempcnta\else
   {\advance\@tempcnta\@ne\ifnum\@tempcnta=\@tempcntb \else \def\@citea{--}\fi
    \advance\@tempcnta\m@ne\the\@tempcnta\@citea\the\@tempcntb}\fi\fi}
 
\makeatother
\begin{titlepage}
\pagenumbering{roman}
\CERNpreprint{\DpPaperGroup}{\DpPaperRef} 
\date{{\small\DpDate}} 
\title{\DpTitle} 
\address{\DpAuthors} 
\begin{shortabs} 
\noindent
\noindent

         $\eta_{c}(2980)$ production in $\gamma\gamma$ interactions
         has been detected via its decays into
         $K^{0}_{s}K^{\pm}\pi^{\mp}$,
         $K^+K^-K^+K^-$ and ${K^+}{K^-}{\pi^+}{\pi^-}$ in the data
         taken with the DELPHI detector at LEP1 and LEP2 energies.
        The  two-photon
         radiative width averaged over all observed decay channels is
          $\Gamma_{\gamma\gamma}$ = 13.9 $\pm$ 2.0  (stat.) $\pm$ 1.4(syst.)
         $\pm$ 2.7 (BR)  keV.
         No direct decay channel
         $\eta_c\rightarrow \pi^+\pi^-\pi^+\pi^-$ has been observed.
         An upper limit  $\Gamma_{\gamma\gamma}$ $<$ 5.5  keV at 95\%
         confidence level  has been evaluated for 
         this decay mode.

\end{shortabs}
\vfill
\begin{center}
\DpSubmit \ \\ 
\DpComment \ \\
\DpEMail \ \\
\end{center}
\vfill
\clearpage
\headsep 10.0pt
\addtolength{\textheight}{10mm}
\addtolength{\footskip}{-5mm}
\begingroup
%
\newcommand{\DpName}[2]{\hbox{#1$^{\ref{#2}}$},\hfill}
\newcommand{\DpNameTwo}[3]{\hbox{#1$^{\ref{#2},\ref{#3}}$},\hfill}
\newcommand{\DpNameThree}[4]{\hbox{#1$^{\ref{#2},\ref{#3},\ref{#4}}$},\hfill}
\newskip\Bigfill \Bigfill = 0pt plus 1000fill
\newcommand{\DpNameLast}[2]{\hbox{#1$^{\ref{#2}}$}\hspace{\Bigfill}}
%
\footnotesize
\noindent
\DpName{J.Abdallah}{LPNHE}
\DpName{P.Abreu}{LIP}
\DpName{W.Adam}{VIENNA}
\DpName{P.Adzic}{DEMOKRITOS}
\DpName{T.Albrecht}{KARLSRUHE}
\DpName{T.Alderweireld}{AIM}
\DpName{R.Alemany-Fernandez}{CERN}
\DpName{T.Allmendinger}{KARLSRUHE}
\DpName{P.P.Allport}{LIVERPOOL}
\DpName{U.Amaldi}{MILANO2}
% ...
\normalsize
\endgroup
\titlefoot{Department of Physics and Astronomy, Iowa State
     University, Ames IA 50011-3160, USA
    \label{AMES}}
\titlefoot{Physics Department, Universiteit Antwerpen,
     Universiteitsplein 1, B-2610 Antwerpen, Belgium \\
     \indent~~and IIHE, ULB-VUB,
     Pleinlaan 2, B-1050 Brussels, Belgium \\
     \indent~~and Facult\'e des Sciences,
     Univ. de l'Etat Mons, Av. Maistriau 19, B-7000 Mons, Belgium
    \label{AIM}}
\titlefoot{Physics Laboratory, University of Athens, Solonos Str.
     104, GR-10680 Athens, Greece
    \label{ATHENS}}
\titlefoot{Department of Physics, University of Bergen,
     All\'egaten 55, NO-5007 Bergen, Norway
    \label{BERGEN}}
\titlefoot{Dipartimento di Fisica, Universit\`a di Bologna and INFN,
     Via Irnerio 46, IT-40126 Bologna, Italy
    \label{BOLOGNA}}
\titlefoot{Centro Brasileiro de Pesquisas F\'{\i}sicas, rua Xavier Sigaud 150,
     BR-22290 Rio de Janeiro, Brazil \\
     \indent~~and Depto. de F\'{\i}sica, Pont. Univ. Cat\'olica,
     C.P. 38071 BR-22453 Rio de Janeiro, Brazil \\
     \indent~~and Inst. de F\'{\i}sica, Univ. Estadual do Rio de Janeiro,
     rua S\~{a}o Francisco Xavier 524, Rio de Janeiro, Brazil
    \label{BRASIL}}
\titlefoot{Coll\`ege de France, Lab. de Physique Corpusculaire, IN2P3-CNRS,
     FR-75231 Paris Cedex 05, France
    \label{CDF}}
\titlefoot{CERN, CH-1211 Geneva 23, Switzerland
    \label{CERN}}
\titlefoot{Institut de Recherches Subatomiques, IN2P3 - CNRS/ULP - BP20,
     FR-67037 Strasbourg Cedex, France
    \label{CRN}}
\titlefoot{Now at DESY-Zeuthen, Platanenallee 6, D-15735 Zeuthen, Germany
    \label{DESY}}
\titlefoot{Institute of Nuclear Physics, N.C.S.R. Demokritos,
     P.O. Box 60228, GR-15310 Athens, Greece
    \label{DEMOKRITOS}}
\titlefoot{FZU, Inst. of Phys. of the C.A.S. High Energy Physics Division,
     Na Slovance 2, CZ-180 40, Praha 8, Czech Republic
    \label{FZU}}
\titlefoot{Dipartimento di Fisica, Universit\`a di Genova and INFN,
     Via Dodecaneso 33, IT-16146 Genova, Italy
    \label{GENOVA}}
\titlefoot{Institut des Sciences Nucl\'eaires, IN2P3-CNRS, Universit\'e
     de Grenoble 1, FR-38026 Grenoble Cedex, France
    \label{GRENOBLE}}
\titlefoot{Helsinki Institute of Physics, P.O. Box 64,
     FIN-00014 University of Helsinki, Finland
    \label{HELSINKI}}
\titlefoot{Joint Institute for Nuclear Research, Dubna, Head Post
     Office, P.O. Box 79, RU-101 000 Moscow, Russian Federation
    \label{JINR}}
\titlefoot{Institut f\"ur Experimentelle Kernphysik,
     Universit\"at Karlsruhe, Postfach 6980, DE-76128 Karlsruhe,
     Germany
    \label{KARLSRUHE}}
\titlefoot{Institute of Nuclear Physics,Ul. Kawiory 26a,
     PL-30055 Krakow, Poland
    \label{KRAKOW1}}
\titlefoot{Faculty of Physics and Nuclear Techniques, University of Mining
     and Metallurgy, PL-30055 Krakow, Poland
    \label{KRAKOW2}}
\titlefoot{Universit\'e de Paris-Sud, Lab. de l'Acc\'el\'erateur
     Lin\'eaire, IN2P3-CNRS, B\^{a}t. 200, FR-91405 Orsay Cedex, France
    \label{LAL}}
\titlefoot{School of Physics and Chemistry, University of Lancaster,
     Lancaster LA1 4YB, UK
    \label{LANCASTER}}
\titlefoot{LIP, IST, FCUL - Av. Elias Garcia, 14-$1^{o}$,
     PT-1000 Lisboa Codex, Portugal
    \label{LIP}}
\titlefoot{Department of Physics, University of Liverpool, P.O.
     Box 147, Liverpool L69 3BX, UK
    \label{LIVERPOOL}}
\titlefoot{Dept. of Physics and Astronomy, Kelvin Building,
     University of Glasgow, Glasgow G12 8QQ
    \label{GLASGOW}}
\titlefoot{LPNHE, IN2P3-CNRS, Univ.~Paris VI et VII, Tour 33 (RdC),
     4 place Jussieu, FR-75252 Paris Cedex 05, France
    \label{LPNHE}}
\titlefoot{Department of Physics, University of Lund,
     S\"olvegatan 14, SE-223 63 Lund, Sweden
    \label{LUND}}
\titlefoot{Universit\'e Claude Bernard de Lyon, IPNL, IN2P3-CNRS,
     FR-69622 Villeurbanne Cedex, France
    \label{LYON}}
\titlefoot{Dipartimento di Fisica, Universit\`a di Milano and INFN-MILANO,
     Via Celoria 16, IT-20133 Milan, Italy
    \label{MILANO}}
\titlefoot{Dipartimento di Fisica, Univ. di Milano-Bicocca and
     INFN-MILANO, Piazza della Scienza 2, IT-20126 Milan, Italy
    \label{MILANO2}}
\titlefoot{IPNP of MFF, Charles Univ., Areal MFF,
     V Holesovickach 2, CZ-180 00, Praha 8, Czech Republic
    \label{NC}}
\titlefoot{NIKHEF, Postbus 41882, NL-1009 DB
     Amsterdam, The Netherlands
    \label{NIKHEF}}
\titlefoot{National Technical University, Physics Department,
     Zografou Campus, GR-15773 Athens, Greece
    \label{NTU-ATHENS}}
\titlefoot{Physics Department, University of Oslo, Blindern,
     NO-0316 Oslo, Norway
    \label{OSLO}}
\titlefoot{Dpto. Fisica, Univ. Oviedo, Avda. Calvo Sotelo
     s/n, ES-33007 Oviedo, Spain
    \label{OVIEDO}}
\titlefoot{Department of Physics, University of Oxford,
     Keble Road, Oxford OX1 3RH, UK
    \label{OXFORD}}
\titlefoot{Dipartimento di Fisica, Universit\`a di Padova and
     INFN, Via Marzolo 8, IT-35131 Padua, Italy
    \label{PADOVA}}
\titlefoot{Rutherford Appleton Laboratory, Chilton, Didcot
     OX11 OQX, UK
    \label{RAL}}
\titlefoot{Dipartimento di Fisica, Universit\`a di Roma II and
     INFN, Tor Vergata, IT-00173 Rome, Italy
    \label{ROMA2}}
\titlefoot{Dipartimento di Fisica, Universit\`a di Roma III and
     INFN, Via della Vasca Navale 84, IT-00146 Rome, Italy
    \label{ROMA3}}
\titlefoot{DAPNIA/Service de Physique des Particules,
     CEA-Saclay, FR-91191 Gif-sur-Yvette Cedex, France
    \label{SACLAY}}
\titlefoot{Instituto de Fisica de Cantabria (CSIC-UC), Avda.
     los Castros s/n, ES-39006 Santander, Spain
    \label{SANTANDER}}
\titlefoot{Inst. for High Energy Physics, Serpukov
     P.O. Box 35, Protvino, (Moscow Region), Russian Federation
    \label{SERPUKHOV}}
\titlefoot{J. Stefan Institute, Jamova 39, SI-1000 Ljubljana, Slovenia
     and Laboratory for Astroparticle Physics,\\
     \indent~~Nova Gorica Polytechnic, Kostanjeviska 16a, SI-5000 Nova Gorica, Slovenia, \\
     \indent~~and Department of Physics, University of Ljubljana,
     SI-1000 Ljubljana, Slovenia
    \label{SLOVENIJA}}
\titlefoot{Fysikum, Stockholm University,
     Box 6730, SE-113 85 Stockholm, Sweden
    \label{STOCKHOLM}}
\titlefoot{Dipartimento di Fisica Sperimentale, Universit\`a di
     Torino and INFN, Via P. Giuria 1, IT-10125 Turin, Italy
    \label{TORINO}}
\titlefoot{INFN,Sezione di Torino, and Dipartimento di Fisica Teorica,
     Universit\`a di Torino, Via P. Giuria 1,\\
     \indent~~IT-10125 Turin, Italy
    \label{TORINOTH}}
\titlefoot{Dipartimento di Fisica, Universit\`a di Trieste and
     INFN, Via A. Valerio 2, IT-34127 Trieste, Italy \\
     \indent~~and Istituto di Fisica, Universit\`a di Udine,
     IT-33100 Udine, Italy
    \label{TU}}
\titlefoot{Univ. Federal do Rio de Janeiro, C.P. 68528
     Cidade Univ., Ilha do Fund\~ao
     BR-21945-970 Rio de Janeiro, Brazil
    \label{UFRJ}}
\titlefoot{Department of Radiation Sciences, University of
     Uppsala, P.O. Box 535, SE-751 21 Uppsala, Sweden
    \label{UPPSALA}}
\titlefoot{IFIC, Valencia-CSIC, and D.F.A.M.N., U. de Valencia,
     Avda. Dr. Moliner 50, ES-46100 Burjassot (Valencia), Spain
    \label{VALENCIA}}
\titlefoot{Institut f\"ur Hochenergiephysik, \"Osterr. Akad.
     d. Wissensch., Nikolsdorfergasse 18, AT-1050 Vienna, Austria
    \label{VIENNA}}
\titlefoot{Inst. Nuclear Studies and University of Warsaw, Ul.
     Hoza 69, PL-00681 Warsaw, Poland
    \label{WARSZAWA}}
\titlefoot{Fachbereich Physik, University of Wuppertal, Postfach
     100 127, DE-42097 Wuppertal, Germany \\
\noindent
{$^\dagger$~deceased}
    \label{WUPPERTAL}}
\addtolength{\textheight}{-10mm}
\addtolength{\footskip}{5mm}
\clearpage
\headsep 30.0pt
\end{titlepage}
%
\pagenumbering{arabic} 
\setcounter{footnote}{0} %
\large
\section{Introduction}

Among $\gamma \gamma$ induced final states, those with exclusive meson
resonance production play an important role, since the measurement of the
production cross-section and the corresponding  radiative width  provide
information on the quark-gluon structure of the investigated particle.
Among these final states, those with mesons built up
of heavy quarks are particularly interesting since
such mesons can be described with nonrelativistic models.
In particular, a precise measurement of the
two-photon partial width for charmonium states would provide valuable
information on QCD corrections to $c\bar{c}$ quarkonium.

  The very first estimations of the $\eta_c$ partial width,
$\Gamma_{\gamma\gamma}(\eta_c)$, were obtained from its ratios to the known
widths for $\psi \rightarrow \mu^+\mu^-$ and $\eta_c \rightarrow gg$
giving values of 8 keV and 4 keV respectively~\cite{kwong}.
Different models and corrections were applied to them
later, giving values from 3 to 14 keV, see~\cite{rein} and
references therein. An even bigger discrepancy
is observed between values obtained by numerous experimental groups.
Among them there are many experiments where two interacting
photons radiated by electron and positron beams
couple to this resonant state,~\cite{PLUT}-\cite{CLEO2}. The results for
$\Gamma_{\gamma\gamma}(\eta_c)$   range from 4 keV to 27 keV.

In this paper we report on the production and decays of the $\eta_c$
resonance using data collected by the DELPHI detector during the  period
1994-1999 corresponding to a range of centre-of-mass energies from 90 GeV up to
202 GeV and an integrated luminosity of ${\cal L}=531$ pb$^{-1}$. The aim of this
analysis was to determine the radiative width of the $\eta_c$ resonance
 separately for each  decay channel,
using the production process:\\
\begin{equation}
 e^{+}e^{-} \rightarrow e^+e^-\eta_c(2980)
\end{equation}
 on four-body final states where a distinct signal of the $\eta_c$
 resonance has been observed.
To increase
the sensitivity for $\eta_c$ production, we do not require information
on the polar angle of the scattered electrons (no tag mode). The superiority
of LEP with respect to previous experiments is the higher energy and
resulting higher production cross-section for this reaction.

We have analysed the following exclusive final states:
\begin{equation}
       \eta_c \rightarrow K^{0}_{s} K^{\pm} \pi^{\mp}
\end{equation}
\begin{equation}
       \eta_c \rightarrow K^+ K^- K^+ K^-
\end{equation}
\begin{equation}
       \eta_c \rightarrow K^+ K^- \pi^+ \pi^-
\end{equation}
\begin{equation}
       \eta_c \rightarrow \pi^+ \pi^- \pi^+ \pi^-
\end{equation}

\section{Detector}

A general description of the DELPHI detector can be found elsewhere
~\cite{detdel}. The main features relevant to this analysis are particle
tracking and identification. Due to the low momenta of the
decay products, their identification  is  based on measurement of
ionization losses (dE/dx) in the Time Projection Chamber (TPC). 
The particle momenta are determined from track
reconstruction and make use of the Vertex Detector, the Inner and Outer
Detectors and the TPC. The tracks with lower polar angles are reconstructed
in Forward Chambers A/B.

The single track trigger efficiency, expressed in terms of transverse track
momentum, has an influence on the overall efficiency of final states
produced in $\gamma \gamma$ collisions where the hadrons have rather low
momenta. Having four particles in the final state, originating from the decay
 of a relatively heavy $\eta_c$(2980) resonance,
results in a large trigger efficiency for an event according to the formula:
\begin{equation}
{\cal E}_{ev}= 1-(1-\epsilon_1)\times (1-\epsilon_2)\times (1-\epsilon_3)
\times (1-\epsilon_4)
\end{equation}
where ${\cal E}_{ev}$ is the total trigger efficiency for an event and
the $\epsilon_i$ is  the  single track efficiency which depends on the 
transverse momentum. A brief description of the trigger system is presented in
~\cite{trig,canal}.\\


\section{General Data Selection}

Data were taken only from running periods
when the TPC
was fully operational thus ensuring good particle identification. There was no
requirement on detecting either scattered electron.
Candidates for the $\eta_c$(2980) decay channels (2)-(5)
 were selected by requiring:
\begin{itemize}
\item exactly four charged particle tracks with zero total charge,
          coming from the primary interaction region or two tracks originating from the
          primary vertex and two tracks originating from a secondary vertex,
\item the track impact parameters measured with respect to the
          z-axis (beam axis) to be smaller than 10 cm and those measured in
          the plane perpendicular to the z-axis smaller than 4 cm,
\item the momentum of each particle to be larger than 0.1 GeV/c,
\item the square of the total  transverse momentum, $(\Sigma \vec{p_t})^2$, of
          charged particles to be less than 1.0 (GeV/c)$^2$,
\item each track to pass through the TPC,
\item the total detected energy of charged particles to be less than 10 GeV,
\item no particles identified as electrons or muons by
          the standard lepton identification algorithms,
\item the track lengths to be longer than 30 cm,
\item the total energy deposit in the electromagnetic calorimeter from
          neutral particles to be less than 3 GeV,
\item the charged particles to have polar angles between 20$^\circ$ and
          160$^\circ$.
 \end{itemize}
Additional criteria which are specific to particular channels are discussed
in the next section.

All experimental requirements used in the
analyses presented below were chosen to be the same for all data sets
corresponding to various beam energies.

\section{Analysis}

\begin{figure}[H]
\centering
\epsfig{figure=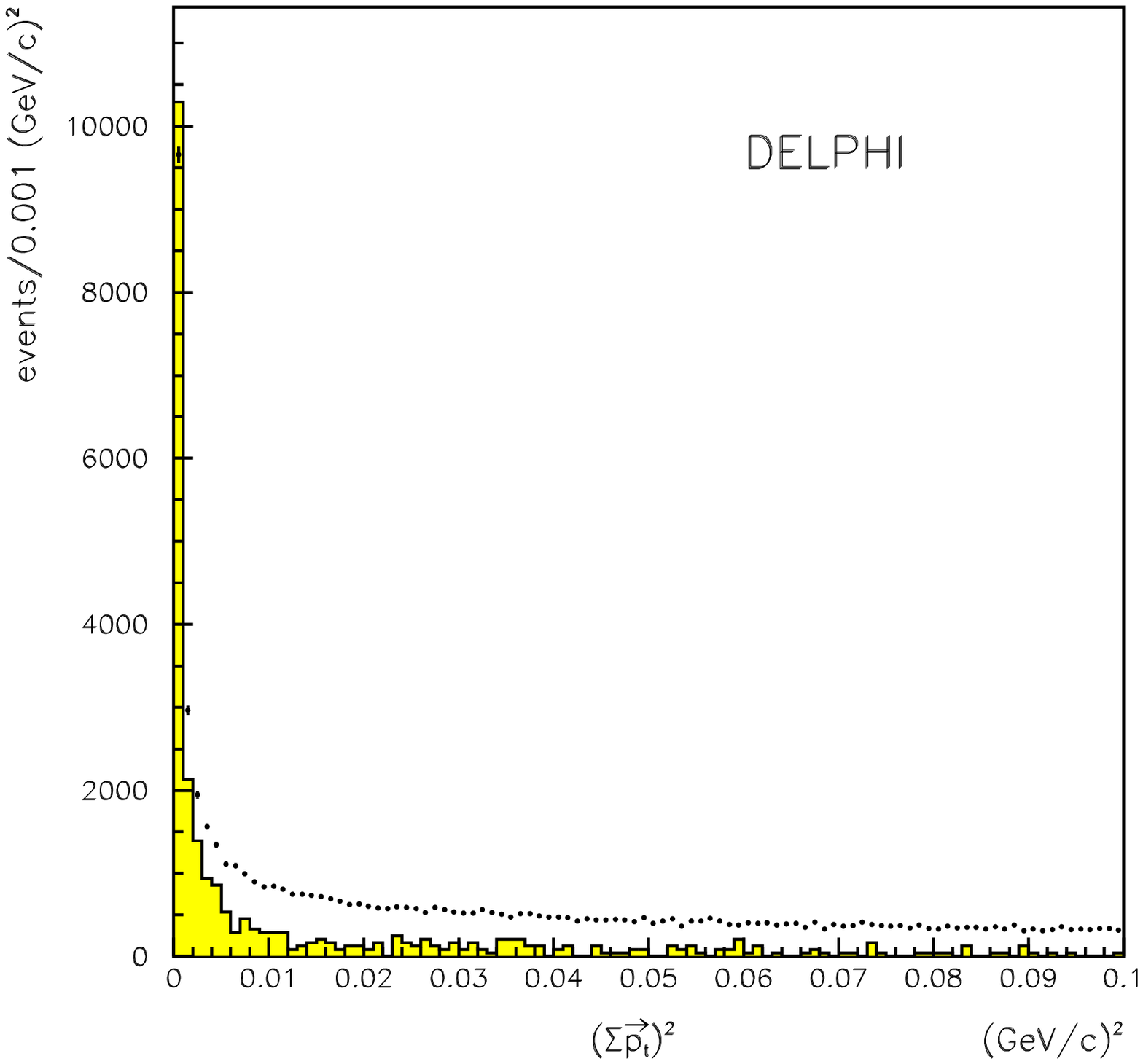,height=9cm}
\caption{ The square of the total transverse momentum of the hadronic system.
Points represent the real data sample after the general data selection.
 The histogram shows this distribution for 
dedicated  $\eta_c$ production simulation sample with an arbitrary
normalization.}
\label{f:pt}
\end{figure}

In $\gamma \gamma$  events almost all the available energy and momentum is
carried away by the electron and positron which are scattered at very small
angles. Therefore the $(\Sigma \vec{p_t})^2$ distribution of the hadronic
 system is
peaked at low values, as shown in Fig.\ref{f:pt}.
 To suppress  background events
which do not
originate from $\gamma \gamma$ collisions, the total transverse momentum
squared  of hadrons in
the exclusive process (1) should  be smaller than 0.04 (GeV/c)$^2$.

In order to calculate the acceptance and detection efficiency,
a Monte Carlo generation
program has been used, with the full kinematics of a system produced in
$\gamma \gamma$ interactions. All kinematical variables necessary for the
description of the two-photon processes were generated using algorithms
taken from the package described in~\cite{Verm}. The matrix element,
factorized into the flux of quasi-real transverse photons and a 
covariant amplitude
describing both the two-photon $\eta_c$ production and its decay, has been
implemented~\cite{mat}. For a better understanding of the
$\eta_c$  four pion decay mode we have also determined the efficiency for
$\eta_c \rightarrow \rho^0\rho^0 \rightarrow{\pi}^+ {\pi}^- {\pi}^+ {\pi}^-$
with a specific symmetrized matrix element~\cite{mat}. The Monte Carlo
generated events were passed through the standard DELPHI detector simulation
procedure \cite{detdel}.

 An additional
factor contributing to the overall efficiency comes from the trigger acceptance.
The trigger simulation following the cuts used for $\eta_c$ selection in the
real data has been applied to events after detector simulation. 
An event was accepted according to a weight calculated on the basis of the
single track efficiency, parameterized as a function of the transverse
momentum, $p_t$,  and ranges from 20$\%$ for $p_t$= 0.5
GeV/c to about 95$\%$ at $p_t$= 2 GeV/c~\cite{trig,canal}. Owing to the
relatively large mass of the $\eta_c$ resonant state, the overall trigger
efficiency per event was about 90$\%$ for channels with pions and about
85$\%$ for the $K^+K^-K^+K^-$ final state.

\begin{figure}[H]
\centering
\epsfig{figure=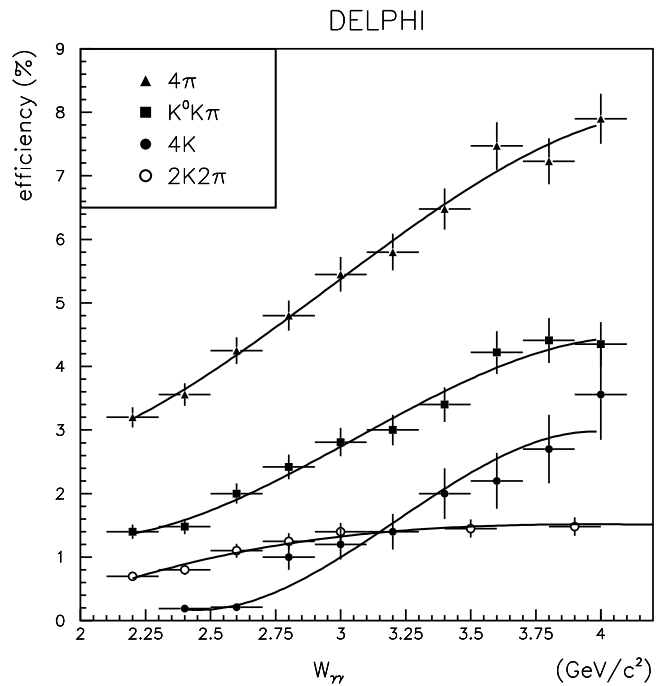,height=9cm}
\caption{ The average luminosity-weighted efficiences for different $\eta_c$
decay final states as a function of the corresponding
invariant mass. In the decay to $K^0$, its branching fraction to
$\pi^+\pi^-$ has been taken into account.}
\label{f:eff}
\end{figure}

The total efficiency was calculated bin-by-bin in invariant mass
 by comparing the generated
invariant mass distribution with that obtained from the detector simulation
after the selection cuts and trigger acceptance. The efficiency 
for each decay mode as a
function of the invariant mass is shown in Fig.\ref{f:eff}.
It should be noted that particle identification was essential for all the
channels analysed and was based on 
dE/dx energy loss measurements~\cite{detdel}.

\subsection{$\eta_c \rightarrow K^{0}_{s} K^{\pm} \pi^{\mp}$}
\label{sub:k0kpi}
\begin{figure}[h]
\centering
\epsfig{figure=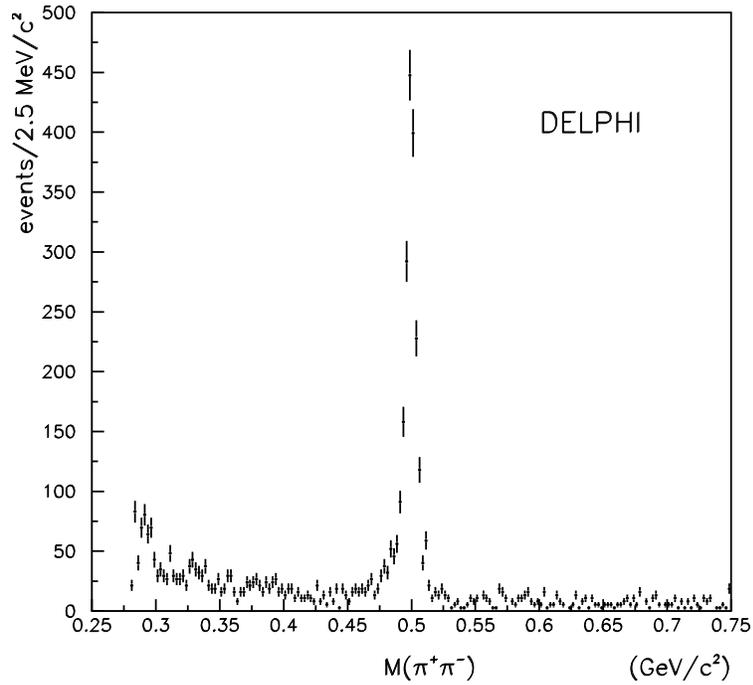,height=9cm}
\caption{ Invariant mass of two particles originating from a secondary
vertex (summed over all energy samples).}
\label{f:k0}
\end{figure}

For the decay chain $\eta_c \rightarrow K^{0}_{s} K^{\pm} \pi^{\mp}
\rightarrow \pi^+\pi^-  K^{\pm} \pi^{\mp}$  the $K^{0}_{s} \rightarrow {\pi}^+ {\pi}^-$ decay is identified by
taking advantage of the relatively large  $K^{0}_{s}$ decay length (c$\tau$ =
2.68 cm). Therefore, candidates for this decay mode had to have one
secondary vertex reconstructed using an algorithm which takes pairs of
oppositely charged particle tracks, intersecting them and determining a
secondary vertex. Both momenta are recalculated with respect to the new
decay vertex and an invariant mass is computed. The resulting
$K^{0}_{s}$ candidate mass distribution is shown in Fig.\ref{f:k0}, where clear
evidence of a $K^{0}_{s}$ signal is seen. Only events with an invariant mass
of the two pion candidates, originating from the secondary vertex, in the
range from 0.45 GeV/c$^2$ to 0.55 GeV/c$^2$  have been taken for further
analysis. Of the other two particles which originate from primary interaction
region,  one is identified as a kaon in 80\% of the events selected
with one secondary vertex. Hence the crucial criterion for this decay final
state selection is the reconstruction of the $K^{0}_{s}$ decay vertex.

\subsection{$\eta_c \rightarrow K^+ K^- K^+ K^-$}

\begin{figure}[h]
\centering
\epsfig{figure=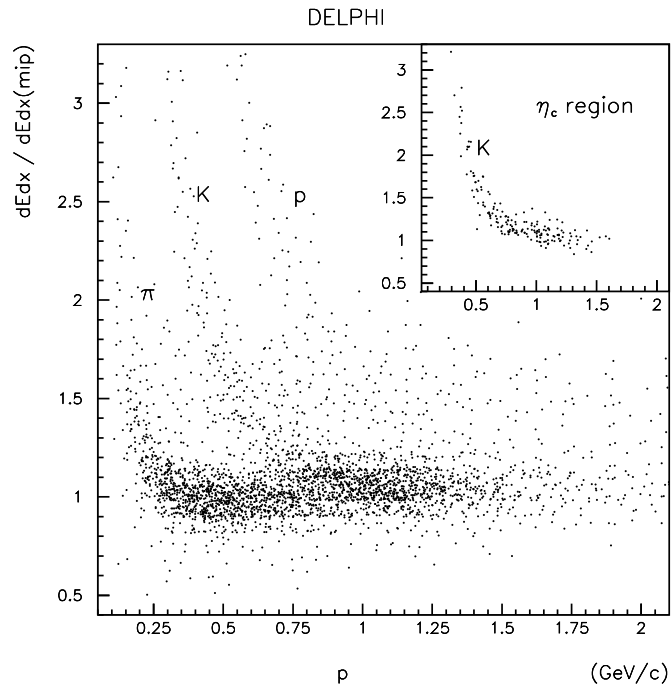,height=9cm}
\caption{ The dE/dx distribution for particles identified as a pion or a kaon
in the data events after the general cuts.
Most of the remaining tracks consist of protons.
 The same distribution for events from
$\eta_c \rightarrow K^+K^-K^+K^-$ is shown in the insert. }
\label{f:dedx}
\end{figure}

Additional requirements for this decay channel are that at least three
particles must be identified as charged kaons and there are
no secondary vertices.
Only kaons with the probability of  identification greater then 0.5 
 were considered.
The dE/dx  distribution for all identified 
 particles  
after the general data selection is plotted in Fig.\ref{f:dedx} with an
 insert for the
distribution of those originating from $\eta_c$ 
(2850~MeV/c$^2$ $<M(K^+ K^- K^+ K^-)<$ 3150~MeV/c$^2$).
For events from this $\eta_c$
mass region,
points originating from the rising part of this distribution unquestionably
correspond to kaons whereas the horizontal part may also contain
pions from  background events and kaons from signal.\\
  A scatter plot (not shown) of the invariant mass of $K^+K^-$
combinations does not indicate any intermediate $\phi\phi$ state. From a fit
to the invariant mass distribution, the number of signal events is estimated
to be about 46.

Since the average particle momentum is particularly low in this channel, a
strong effect could be expected in the invariant mass spectrum resulting
from the single track efficiency of the trigger that might produce a fake signal due
to the small efficiency at threshold. This has been checked on
$\eta_c \rightarrow K^+ K^- K^+ K^-$ events which were generated according
to the $\gamma \gamma$  flux (no resonance shape has been assumed) and then
decayed according to phase-space. These events were then passed through the
trigger and  detector simulations.  No signal resulting from the trigger
activity on the low mass side nor from the experimental cuts on the other
was observed in a region of the invariant masses around 3 GeV/c$^2$,
 corresponding
to the  $\eta_c$
signal. The relatively low
background at the $K^+ K^- K^+ K^-$ invariant mass threshold is explained mostly by the low acceptance and less by the decreased trigger efficiency.
 The trigger
efficiency, as described in previous section turned out to be around 85$\%$ at a mass
of 3 GeV/c$^2$. \\

\subsection{$\eta_c \rightarrow {K^+}{K^-}{\pi^+}{\pi^-}$}

Given the branching ratio for  $\eta_c$ decay into
 ${K^+}{K^-}{\pi^+}{\pi^-}$,
BR=2.0$\pm$0.7\%, and the detector efficiency determined using criteria
 presented below (Fig.\ref{f:eff}),  a significant
signal (of about 4 events per 1 keV of $\eta_c$ radiative width)
 would be expected in this channel.
In order to select these events
 it was required that one of the particles was identified 
as a kaon with probability $\geq$0.5 and two of the three remaining
 particles should satisfy selection criteria for pion identification
with probability $\geq$0.5.
 The identification was
 based  on dE/dx energy losses. 
 All events corresponding to the $K^{0}_{s} K^{\pm} \pi^{\mp}$ signal,
 described in section \ref{sub:k0kpi} have been subtracted
from the selected sample. Since the data sample obtained
may still contain $K^{0}_{s} K^{\pm} \pi^{\mp}$ events with
 no reconstucted secondary vertex,
 the invariant  mass $M_{ik}$ of the two opposite sign particle
combinations (excluding the identified kaon)
 was calculated and events removed if  one of the two $M_{ik}$  masses
 satisfied the condition $|M_{K^{0}_{s}}-M_{ik}|<50$ MeV/c$^2$.
From the data collected by the DELPHI detector during the period
 mentioned in the first section a signal of about 42 events 
is obtained.

The intermediate states of 
$\eta_c \rightarrow {K^+}{K^-}{\pi^+}{\pi^-}$ decay
via one or two $K^{*0}(892)$ have not been observed.

\subsection{$\eta_c \rightarrow {\pi^+}{\pi^-}{\pi^+}{\pi^-}$}

The observation of an $\eta_c \rightarrow {\pi^+}{\pi^-}{\pi^+}{\pi^-}$
decay mode reported by numerous experimental groups remains controversial.
This decay has been found by MARK III~\cite{MARKIII}, DM2~\cite{DM2}, 
TASSO~\cite{TASSO}
 (where the last one did not distinguish between the 4$\pi$
and the $\rho^0 \rho^0$ decay channels). 
Among more recent experiments this
final state has been observed in BES~\cite{bes}. None of the LEP experiments 
  confirm this  decay mode
 providing only an upper limit~\cite{L3P}.
 Good particle identification
is very important since the ${\pi^+}{\pi^-}{\pi^+}{\pi^-}$ final state can
be confused with the ${K^+}{K^-}{\pi^+}{\pi^-}$ decay.

In addition to the general selection and the stringent cut on the total
transverse momentum, it was also required that all particles were pions
with the single track identification  probability $\ge$ 0.5, 
that only one well reconstructed vertex was found and that each track 
had to have at least one hit in the Vertex Detector.
The final selected sample
consists of $\sim$3600 events and shows no enhancement around the nominal mass
of the $\eta_c$ resonance, see Fig.5a.
Using the PDG values~\cite{PDG} for the $\eta_c$ parameters, more
than 60 events would be expected in this channel. An upper limit of 26 
events at 95\% confidence limit has been calculated.
The above, standard selection criteria  lead to an invariant
${\pi^+}{\pi^-}{\pi^+}{\pi^-}$ mass distribution with a large background 
that may shadow the signal.
Further tightening of the total transverse momentum squared cut 
from 0.04 to 0.004 (GeV/c)$^2$  and the identification probability 
from 0.5 to 0.8 
reduces the number of
observed events to about 600   but still the invariant mass distribution
 shows
 no evidence for $\eta_c$, see Fig.5b.
To avoid a selection bias resulting from the low efficiency for the 
 identification of four pions  another selection was performed in which
 only 
three particles
were identified as  pions with probability $\ge$ 0.5, leaving remaining
cuts like in the standard selection, again resulting in no enhancement
 at $\eta_c$
invariant mass region, see Fig.5c. 
 A search for the intermediate decay mode, $\rho^0\rho^0$, through
an analysis of the two-dimensional plot of the invariant mass of one
${\pi^+}{\pi^-}$ system versus that of the remaining ${\pi^+}{\pi^-}$ pair
has been also performed. Events from the $\rho^0\rho^0$ mass window were
selected and used for the calculation of the 
${\pi^+}{\pi^-}{\pi^+}{\pi^-}$
invariant mass spectrum. Since the  $\eta_c$ signal was  not seen,
these events were attributed to non-resonant $\rho^0 \rho^0$
vector mesons production.

\section{Results}

Experimentally one measures directly the invariant mass
($W_{\gamma\gamma}$) distribution of the $\gamma \gamma$ system,
\begin{equation}
\frac 
{\Delta N_{(e^{+}e^{-} \rightarrow  e^{+}e^{-} \eta_c \rightarrow
 e^{+}e^{-}f)}}
 {\Delta W_{\gamma \gamma}}. 
\end{equation}
where f denotes one of the investigated decay modes.
Given the detector efficiency ${\cal E}_f$, 
the  integrated luminosity $\cal L$,
and  flux $ L_{\gamma\gamma}$
 of the two interacting photons parametrized by well known
 equivalent photon approximation formula,
the invariant mass distribution can be converted into 
two-photon cross section multiplied by corresponding branching ratio:
\begin{equation}
\sigma_{\gamma\gamma \rightarrow  \eta_c}
(W_{\gamma\gamma}) \cdot BR(\eta_c \rightarrow f)~=~\frac
{\Delta N_{(e^{+}e^{-} \rightarrow  e^{+}e^{-} \eta_c\rightarrow e^{+}e^{-}f)}} 
{\Delta W_{\gamma \gamma}
~\cdot ~{\cal L}~\cdot {\cal E}_f(W_{\gamma\gamma})~\cdot
 L_{\gamma\gamma}(W_{\gamma\gamma}) }
\end{equation}
The efficiency was calculated dividing bin-by-bin the
simulated invariant mass distribution for events that passed all the
cuts in the mass interval 2.5-4.0 GeV/c$^2$
 by the invariant mass distribution for the  generated events.
It should be noticed that both the $\gamma \gamma$ flux and the invariant
mass efficiency distribution modify
 the background-to-signal ratio measured in
the side-bands of the $\Delta N/ \Delta W_{\gamma \gamma}$  distribution.

In order to determine the value of the $\eta_c$ radiative width, 
the $\gamma \gamma$
invariant mass
cross-section has been fitted to the Breit Wigner distribution of the form
\begin{equation}
BW(\Gamma_{\gamma\gamma},M_{\eta_c},\Gamma_{tot},W_{\gamma\gamma})=8\pi(2J+1)
\displaystyle \frac{\Gamma_{\gamma\gamma}\Gamma_{tot}}
 {(W_{\gamma\gamma}^2-M_{\eta_c}^2)^2+M_{\eta_c}^2\Gamma_{tot}^2}
\end{equation}
 describing the $\eta_c$ production cross-section convoluted with
a Gaussian mass resolution $G(W_{\gamma\gamma},\sigma)$
 together  with a background parametrization
expressed in terms of polynomial function of the third order 
$P_3(W_{\gamma\gamma})$: 
 
\begin{equation}
\sigma_{\gamma\gamma \rightarrow  \eta_c}(W_{\gamma\gamma})=
\left[BW(\Gamma_{\gamma\gamma},M_{\eta_c},\Gamma_{tot},W_{\gamma\gamma})~
+~P_3(W_{\gamma\gamma})\right]\otimes G(W_{\gamma\gamma},\sigma)
\end{equation}

According to eq.(8) and eq.(9) 
the fit determines the  product of the radiative width and the 
branching ratio, the mass of the resonance and the
 experimental mass resolution $\sigma$.
 All these fitted parameters  have been determined
separately for each data sample because some of them explicitly depend
 on the
energy (two-photon flux) and others on the period of the data collection
(efficiency). 
The total width $\Gamma_{tot}$ of the resonace has been fixed to value
obtained by other experiments~\cite{PDG}.

 The width of the mass
resolution distribution obtained from the above fit coincided 
within $\pm10\%$
with that obtained from the simulated sample. 

The final $\sigma_{\gamma \gamma}$ plots are  average
 distributions  from different samples.
The resulting cross-sections multiplied by the corresponding branching 
ratios
 for $\eta_c \rightarrow K^{0}_{s}K^{\pm}\pi^{\mp}$,
$\eta_c \rightarrow K^+K^-K^+K^-$ and 
 $\eta_c \rightarrow K^+K^-\pi^+\pi^- $
are presented in Fig.\ref{f:sigma}.

A major contribution to the systematic uncertainty
originates from the cuts variation (about  44\% of the
total systematic error) and different fit ranges as well as the choice of
binning (about 28\%). Since both the branching ratio and 
 $\Gamma_{\gamma\gamma}$ (see formula above)
cannot be determined simultaneously, we have used branching ratios
and corresponding uncertaintes
obtained by other experiments, summarized in~\cite{PDG} . This sort
of uncertainty also contributed to the total systematic error (in the
amount of 13\%). The remaining part of  systematic error (15\%)
 was due to the uncertaintes of background shape, trigger
efficiency and integrated luminosity value.

In summary, the final values of the $\eta_c$ radiative width
for the three decay channels investigated are presented in 
Table~\ref{res}:
\\

\begin{table}[h]
{\small
\begin{tabular}{|c|c|c|c|}
\hline
final state &  BR ($\eta_c\rightarrow  final$~\cite{PDG}) [\%] & 
N$_{ev}(\eta_c)$ &
            $\Gamma_{\gamma \gamma}$ [keV]\\ \hline
$K^{0}_sK^{\pm}\pi^{\mp}$ & 1.5$\pm$0.4 & 41 &
          13.3 $\pm$ 2.6(stat.) $\pm$ 2.0(syst.) $\pm$ 3.5(BR) \\ \hline
$K^+K^-\pi^+\pi^-$        & 2.0$\pm$0.7 & 42 &
          14.2 $\pm$ 4.9(stat.) $\pm$ 2.9(syst.) $\pm$ 4.9(BR) \\ \hline 
$ K^+K^-K^+K^-$           & 2.1$\pm$1.2 & 46 &
          16.5 $\pm$ 4.3(stat.) $\pm$ 2.7(syst.) $\pm$ 9.4(BR)  \\ \hline 
$\pi^+\pi^-\pi^+\pi^-$    & 1.2$\pm$0.4 & $<$ 26 & $<$5.5  at 95\% confidence\\
\hline

\end{tabular}
\caption{The branching ratios taken from PDG, the number of events and
radiative widths for the particular decay modes.
 The N$_{ev}(\eta_c)$ is the 
number of events selected for region $M_{\eta_c}\pm$ 150 MeV/c$^2$.
In the case of  four pions final state only an upper limit has been
estimated  assuming that 
all events in the $\eta_c$ mass interval are   background.}
\label{res}
}
\end{table}

For the analysed channels, the results quoted above are the averages of the
LEP1 and LEP2 results.
The product $\Gamma_{\gamma \gamma} \cdot$ BR in
   the analysis of the four charged kaon decay channel is in
agreement, within the large errors, with the result of the ARGUS
Collaboration~\cite{ARGUS}, which gives 0.231 $\pm$ 0.090 (stat.) $\pm$
0.023 (syst.) keV.\\
A weighted mean of the radiative width value 
for the first three channels in Table~\ref{res} with weights inversely
proportional to the total error squared has been determined.\\
The result is:

\begin{center}
$\Gamma_{\gamma \gamma}$=
                    13.9 $\pm$ 2.0(stat.)$\pm$1.4(syst.)$\pm$2.7(BR) keV

\end{center}

\input{acknow.tex}


\newpage

\begin{figure}[h]
\centering
\epsfig{figure=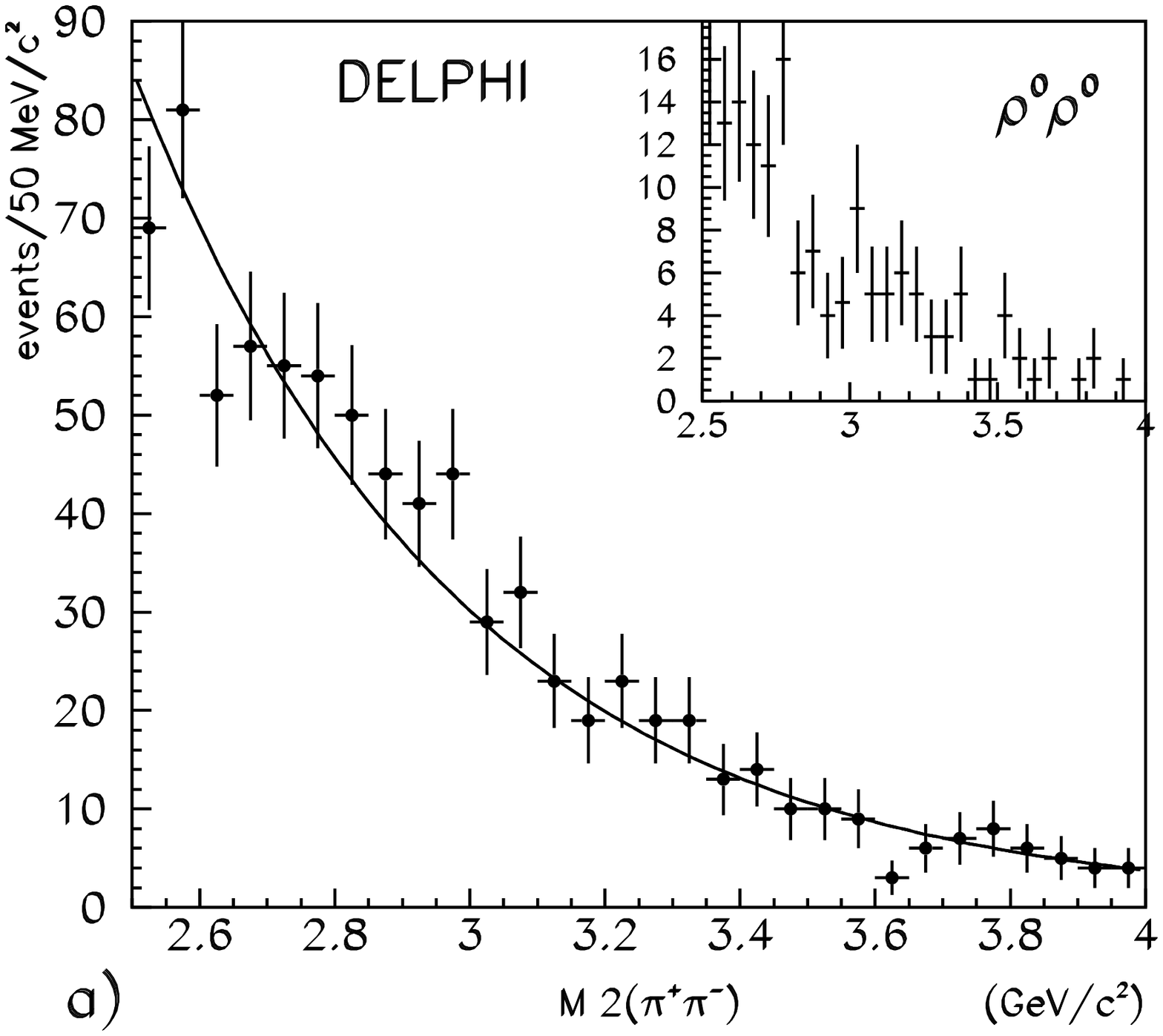,height=7.5cm}\\ \vspace{10pt}
\epsfig{figure=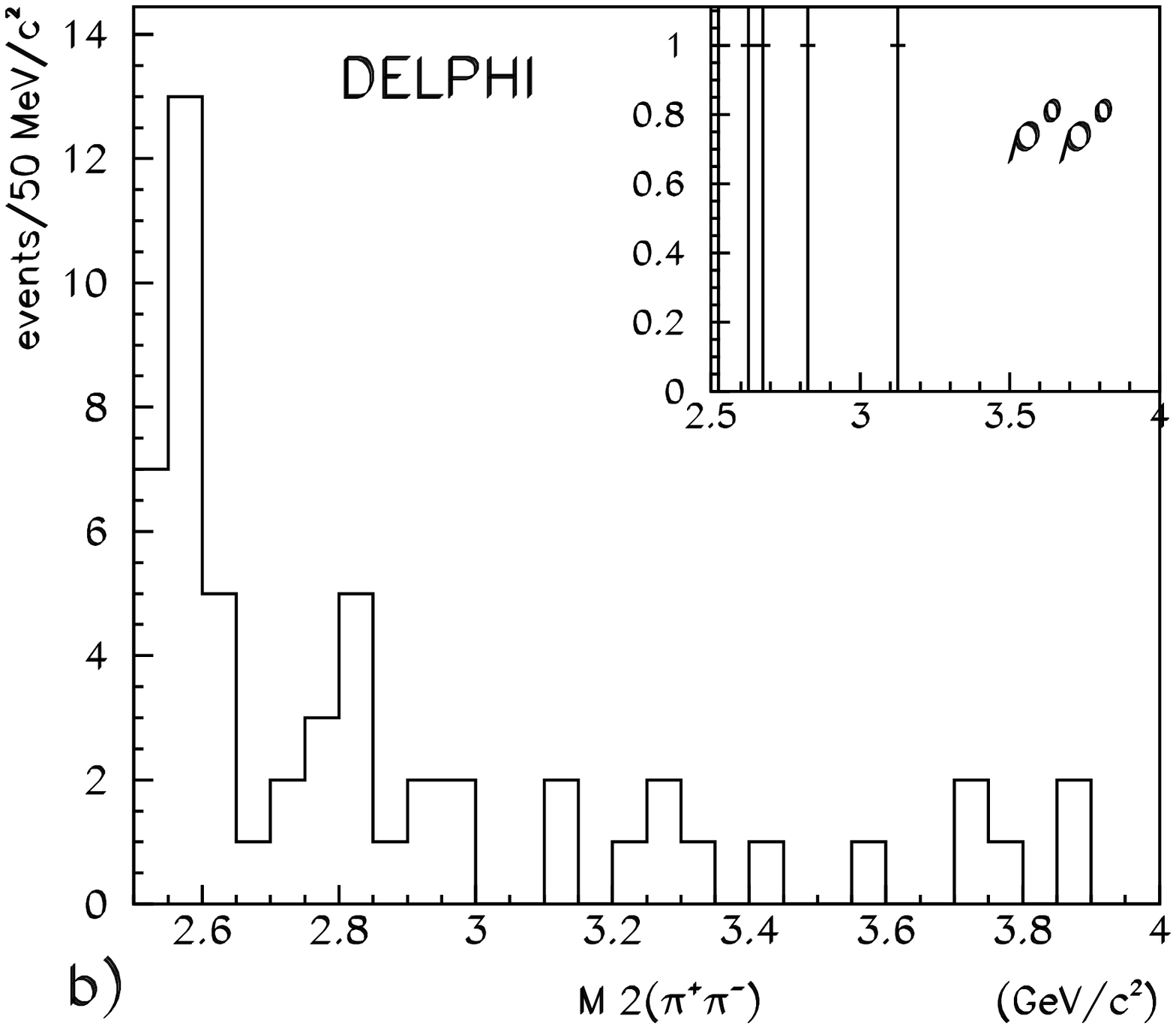,height=7.5cm}\\ \vspace{10pt}
\epsfig{figure=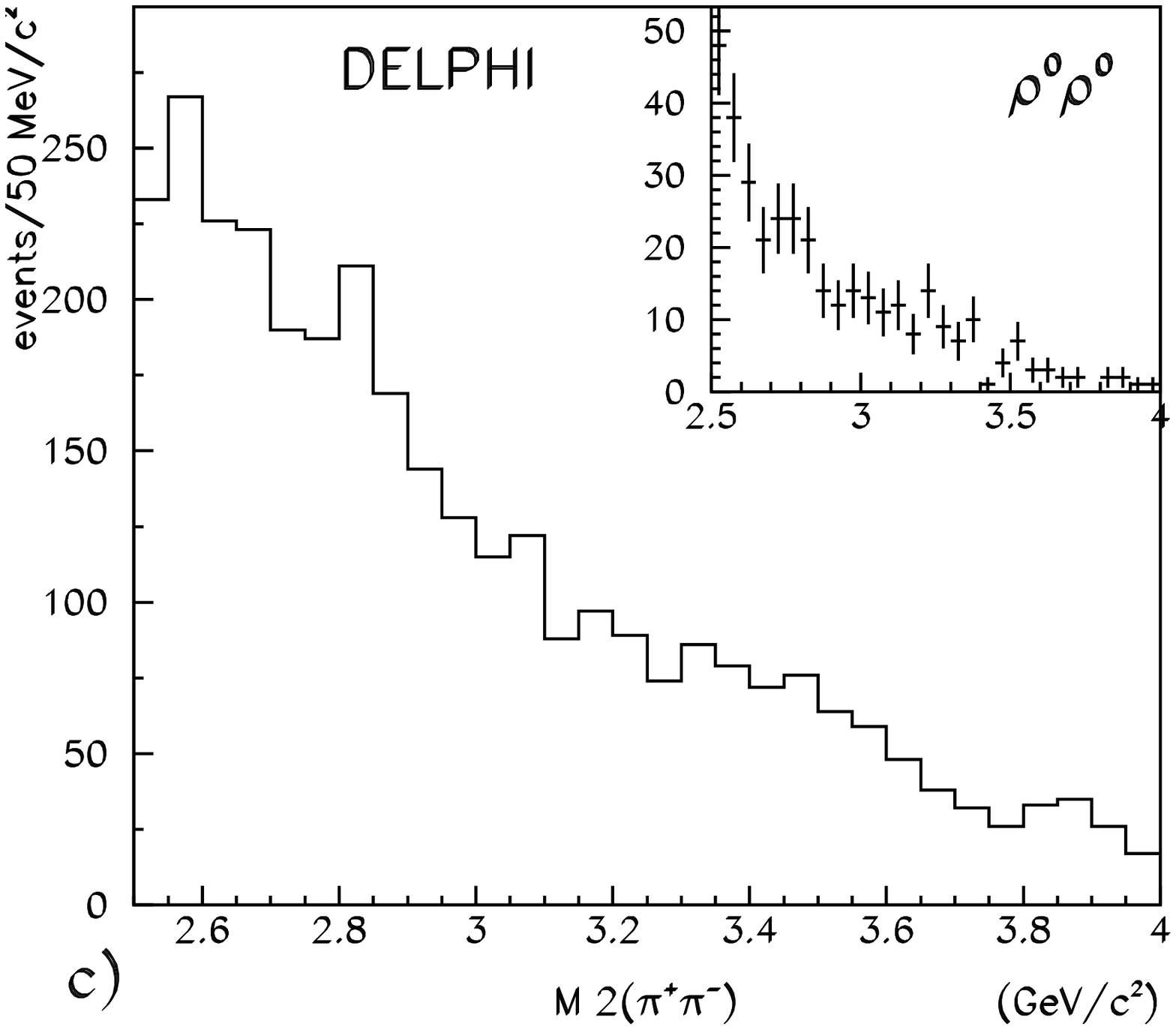,height=7.5cm}
\caption{ Final invariant mass distributions for the $\pi^+\pi^-\pi^+\pi^-$
decay final state. The presented distributions are based on:
 Fig.5a -  the standard selection,
 Fig.5b -  the stringent selection,
 Fig.5c -  the looser selection, all of them
 described in the text.
In the insets the invariant mass distribution of the 
 $\rho^0\rho^0$ events  are also shown.}
\label{f:4pi_ev}
\end{figure}

\begin{figure}[h]
\centering
\epsfig{figure=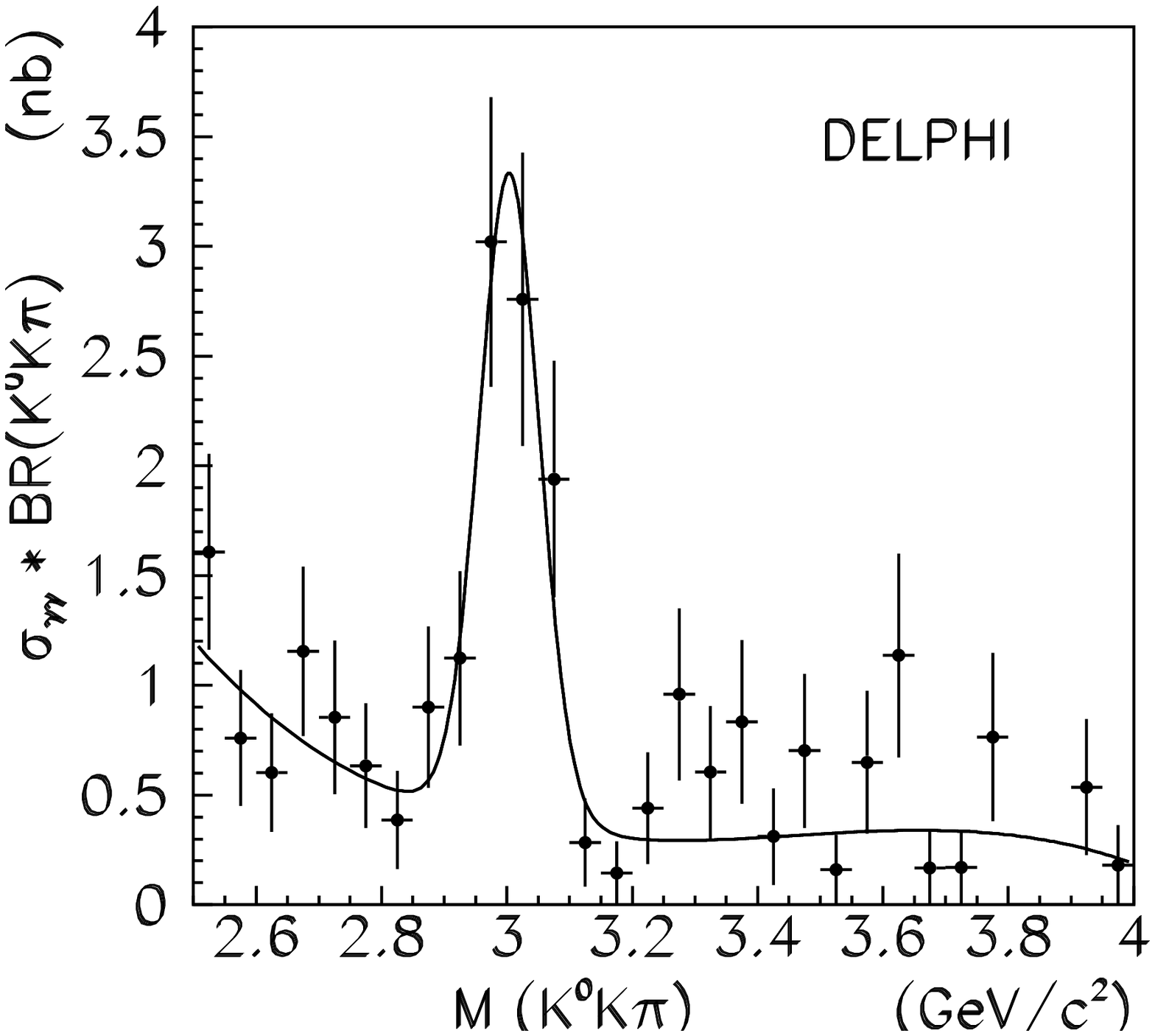,height=7.7cm}\\
~\\
\epsfig{figure=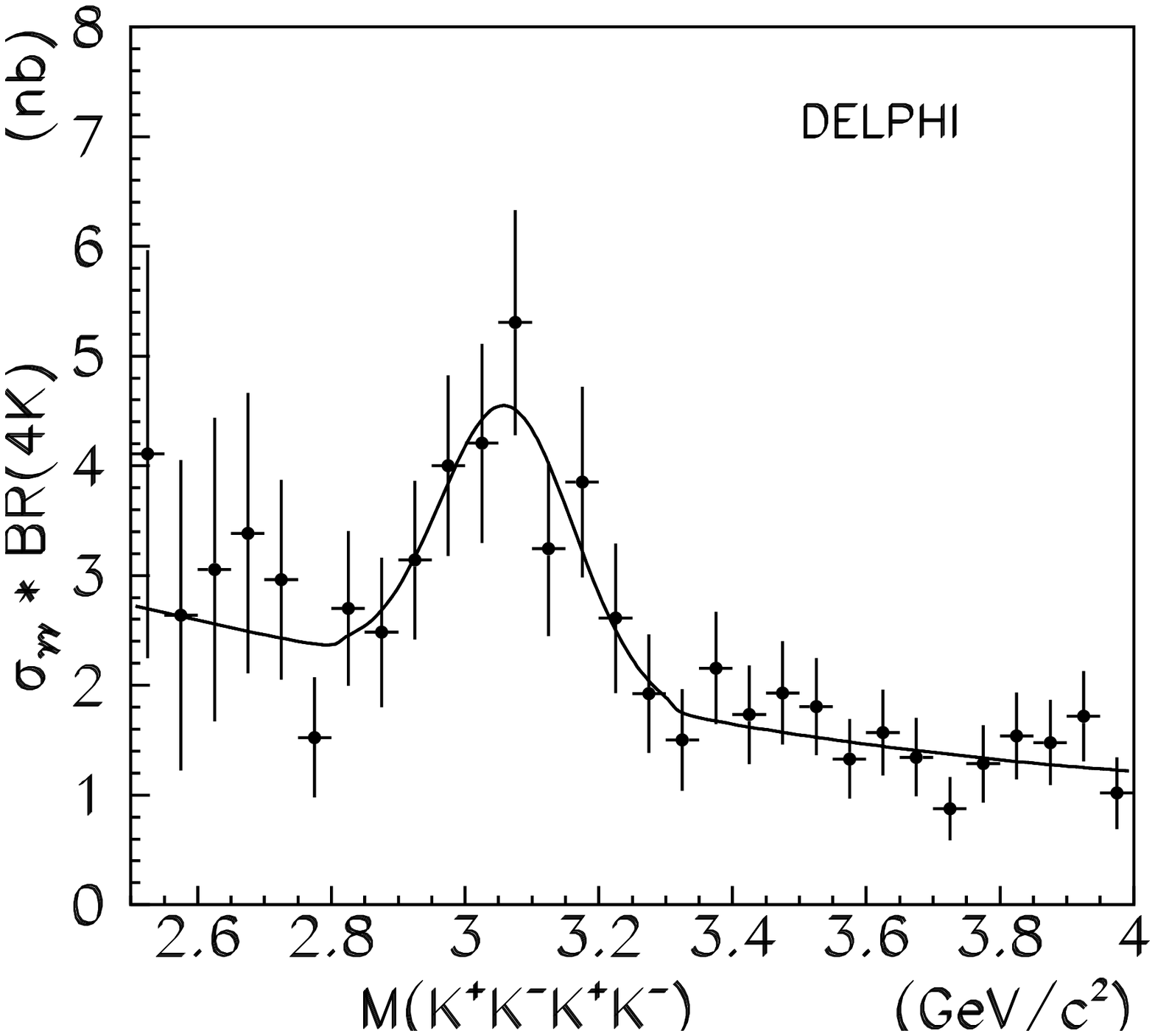,height=7.5cm}\\
~\\
\epsfig{figure=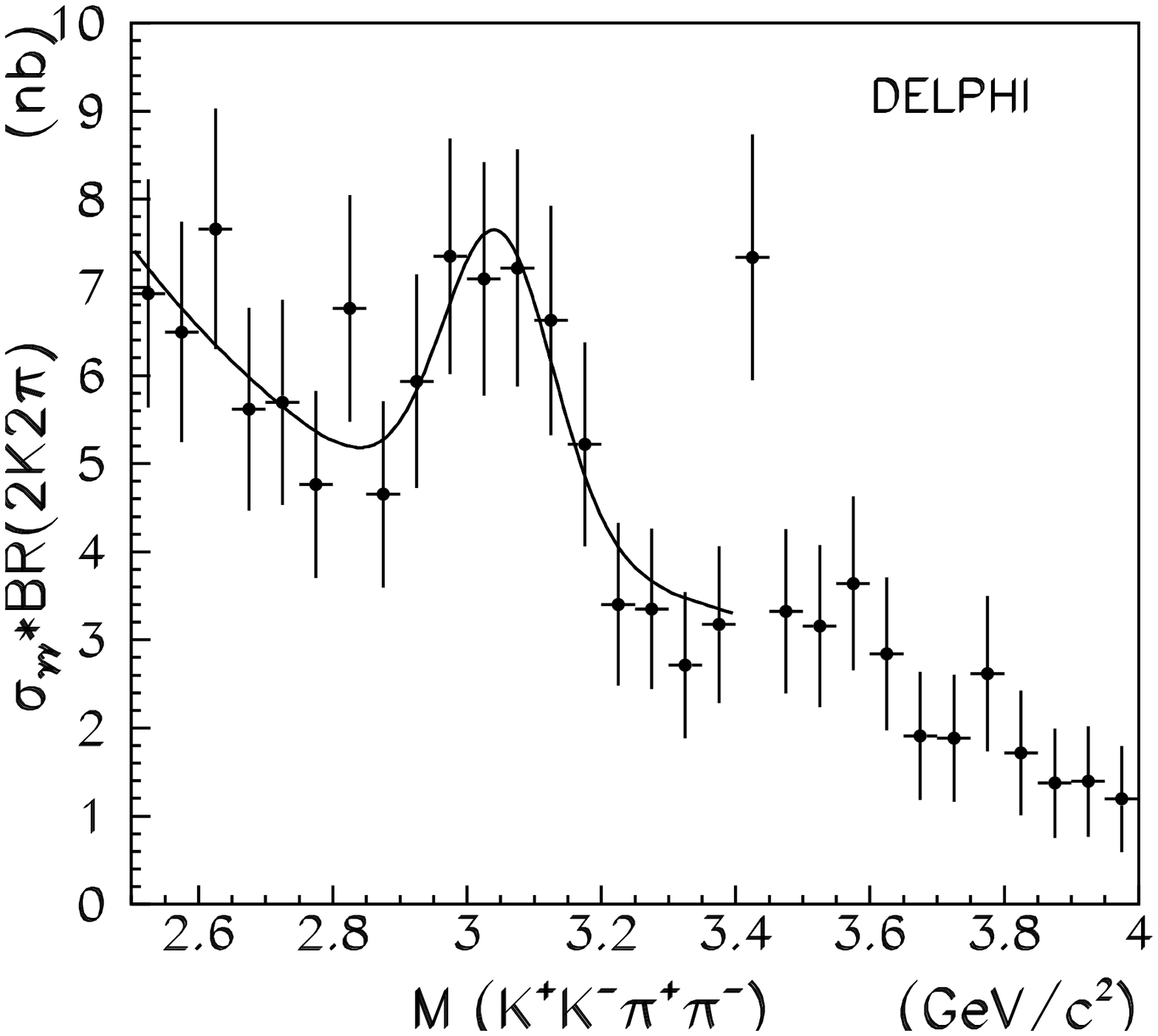,height=7.5cm}
\caption{ $\sigma(\gamma\gamma \rightarrow \eta_c) BR(\eta_c\rightarrow
final)$.
The curve shows the result of the fit described in the text. }
\label{f:sigma}
\end{figure}

\end{document}

%% file: acknow.tex
\subsection*{Acknowledgements}
\vskip 3 mm
 We are greatly indebted to our technical 
collaborators, to the members of the CERN-SL Division for the excellent 
performance of the LEP collider, and to the funding agencies for their
support in building and operating the DELPHI detector.\\
We acknowledge in particular the support of \\
Austrian Federal Ministry of Education, Science and Culture,
GZ 616.364/2-III/2a/98, \\
FNRS--FWO, Flanders Institute to encourage scientific and technological 
research in the industry (IWT), Federal Office for Scientific, Technical
and Cultural affairs (OSTC), Belgium,  \\
FINEP, CNPq, CAPES, FUJB and FAPERJ, Brazil, \\
Czech Ministry of Industry and Trade, GA CR 202/99/1362,\\
Commission of the European Communities (DG XII), \\
Direction des Sciences de la Mati$\grave{\mbox{\rm e}}$re, CEA, France, \\
Bundesministerium f$\ddot{\mbox{\rm u}}$r Bildung, Wissenschaft, Forschung 
und Technologie, Germany,\\
General Secretariat for Research and Technology, Greece, \\
National Science Foundation (NWO) and Foundation for Research on Matter (FOM),
The Netherlands, \\
Norwegian Research Council,  \\
State Committee for Scientific Research, Poland, SPUB-M/CERN/PO3/DZ296/2000,
SPUB-M/CERN/PO3/DZ297/2000 and 2P03B 104 19 and 2P03B 69 23(2002-2004)\\
JNICT--Junta Nacional de Investiga\c{c}\~{a}o Cient\'{\i}fica 
e Tecnol$\acute{\mbox{\rm o}}$gica, Portugal, \\
Vedecka grantova agentura MS SR, Slovakia, Nr. 95/5195/134, \\
Ministry of Science and Technology of the Republic of Slovenia, \\
CICYT, Spain, AEN99-0950 and AEN99-0761,  \\
The Swedish Natural Science Research Council,      \\
Particle Physics and Astronomy Research Council, UK, \\
Department of Energy, USA, DE-FG02-01ER41155, \\
EEC RTN contract HPRN-CT-00292-2002. \\